\documentclass[preprint,number,3p]{elsarticle}
\usepackage{graphicx}
\usepackage{amsmath}
\usepackage{amssymb}
\usepackage{psfrag}
\usepackage{slashed}
\biboptions{sort&compress,numbers}

\def\la{\langle}
\def\ra{\rangle}

\def\A#1#2{\la#1#2\ra}
\def\B#1#2{[#1#2]}
\def\AB#1#2#3{\la#1|#2|#3]}

\def\AA#1#2#3{\la#1|#2|#3\ra}
\def\BB#1#2#3{[#1|#2|#3]}
\def\wh#1{\widehat#1}

\def\sg(#1){\textrm{sign}(#1)}

\def\cF{\mathcal{F}}

\def\tree{\textrm{tree}}
\def\cN{\mathcal{N}}

\def\fl#1{#1^\flat}

\def\qb{\bar{q}}
\def\Sb{\bar{S}}
\def\MHVb{$\overline{\rm MHV}$}

\def\an[#1,#2]{\left\langle#1\,#2\right\rangle}
\def\aq[#1,#2,#3]{\left\langle#1|#2|#3\right]}
\def\qa[#1,#2,#3]{\left[#1|#2|#3\right\rangle}
\def\sq[#1,#2]{\left[#1\,#2\right]}
\def\spa#1.#2{\left\langle#1\,#2\right\rangle}
\def\spab#1.#2.#3{\left\langle#1|#2|#3\right]}
\def\spb#1.#2{\left[#1\,#2\right]}
\def\lor#1.#2{\left(#1\,#2\right)}

\journal{Physics Letters B}
 
\begin{document}

\begin{frontmatter}
\tnotetext[t1]{DESY 10-074, HU-EP 10/16}

\title{Compact QED tree-level amplitudes from dressed BCFW recursion relations}

\author[SB]{Simon D. Badger}
\address[SB]{
Deutches Elektronen-Synchrotron DESY,\\
Platanenallee, 6, D-15738 Zeuthen, Germany}
\ead{simon.badger@desy.de}
\author[JH]{Johannes M. Henn}
\address[JH]{ Institut f\"ur Physik,\\
Humboldt-Universit\"at zu Berlin,\\
Newtonstra\ss{}e15, D-12489 Berlin, Germany}
\ead{henn@physik.hu-berlin.de}

\begin{abstract}
We construct a modified on-shell BCFW recursion relation to derive compact analytic representations
of tree-level amplitudes in QED. 
As an application, we study the amplitudes of a fermion pair coupling
to an arbitrary number of photons and give compact formulae for the NMHV and N$^2$MHV case.
We demonstrate that the new recursion relation reduces the growth in complexity with
additional photons to be exponential rather than factorial.    
\end{abstract}


\end{frontmatter}

\section{Introduction}

Recent years have seen great progress in our understanding of the underlying structures in gauge
theories. On-shell amplitudes are observed to have a much simpler form than their Feynman diagram
representations would suggest. Understanding the origin of these structures allows us to construct
alternative methods which reproduce the simplicity without the need for large intermediate
expressions. On-shell methods like unitarity \cite{Bern:1994zx,Bern:2007dw} and BCFW recursion
\cite{Britto:2004ap,Britto:2005fq} allow us to study multi-particle
and multi-loop amplitudes in a wide range of theories, particularly in theories with a high degree of
super-symmetry. A well studied example of this is $\cN=4$ super Yang-Mills (SYM) where a rich structure of symmetries has
been uncovered in the planar limit \cite{Drummond:2008vq,Drummond:2009fd}. 

Studies of unordered theories, such as gravity, require us to look beyond the planar limit.
Investigations into the the UV properties of perturbative amplitudes $\cN=8$ supergravity
and its connection with $\cN=4$ SYM~\cite{Bern:2002kj,Dixon:2010gz}
have often uncovered new tree-level structures. For recent examples see
\cite{Bern:2008qj,Bern:2010ue,Bern:2010yg,BjerrumBohr:2010zs,Tye:2010dd}.
Analyses of these tree level amplitudes have demonstrated that additional simplifications occur
after one obtains expressions from the ordered case via permutation sums. An example of such
simplifications is an improved behaviour of the $\cN=8$ supergravity tree level amplitudes 
over that of $\cN=4$ SYM amplitudes under large complex deformations of the BCFW shift \cite{ArkaniHamed:tree}.
For explicit expressions of $\cN=8$ supergravity amplitudes see \cite{Bianchi:2008pu,Drummond:2009ge};
(\cite{Drummond:2009ge} was obtained by solving the BCFW recursion
relations in a way similar to $\cN=4$ SYM \cite{Drummond:2008cr})
The latter can be used, through unitarity, to demonstrate the vanishing of triangle coefficients at
one loop \cite{Bern:2007xj,ArkaniHamed:2008gz}.
A result that has also been obtained using a string
based approach \cite{BjerrumBohr:2008ji}.
An interesting spin-off of this no-triangle property in $\cN=8$ super-gravity was the observation that
similar cancellations persist in another gauge theory, this time without super-symmetry, QED \cite{Badger:2008rn}. 

It is also interesting that in such theories the standard on-shell BCFW recursion 
does not yield the most compact representation of the amplitude. In this letter we describe how to use the new information
characterising additional gauge cancellations to construct a modified recursion relation with fewer
number of terms.

Our approach involves a modification to the BCFW recursion relation by changing the form of the
integration kernel of the
contour integral. In the context of gravity amplitudes Spradlin, Volovich and Wen constructed a
recursive system which led to compact expression \cite{Spradlin:bonus}. This system can be
interpreted as adding a single propagator term into the Cauchy integral. Similar modifications have
also been considered recently in the context of boundary terms in BCFW recursion \cite{Feng:2010ku} and
a re-examination of the U(1) decoupling relation \cite{Feng:2010my}. In this paper we extend these
ideas to the case where multiple propagator factors can be used to modify the recursion relation
without introducing a boundary term.

The upshot is that where applicable, the method we propose yields more
compact results compared to those computed via the standard BCFW recursion,
and in addition the large $z$ behaviour of the amplitudes becomes manifest.

Our paper is organised as follows. Firstly we review the Feynman representation of the tree-level
QED amplitudes we will study
and their improved scaling behaviour under the BCFW complex momentum shift. 
We then describe the construction of the dressed recursion relation 
that absorbs certain BCFW diagrams into a modified propagator.
In section \ref{sec:KS} we re-derive the compact MHV formula of Kleiss and
Stirling \cite{Kleiss:qed} using the modified recursion. We then derive new compact formulae for the NMHV and N$^2$MHV
amplitudes and study their improved combinatoric behaviour. In section \ref{sec:mscalar} we demonstrate the
method applies equally 
to amplitudes with a massive scalar before we reach our conclusions.

\section{Tree-Level QED Amplitudes}

In this section we review the tree-level amplitudes of a fermion or massive scalar pair coupling to an arbitrary number
of photons and summarise their behaviour under BCFW shifts.

\subsection{The $q\bar{q}+n(\gamma)$ amplitudes}

The main object we will study in this paper are the well known tree-level amplitudes with a fermion
pair coupling to an arbitrary number of photons,
\begin{align}
  q^{-}(k_q) + \bar{q}^{+}(k_{\qb}) + \gamma^{h_{1}}(k_{1}) + \ldots  \gamma^{h_{n}}(k_{n}) \rightarrow 0 \,, 
\end{align}
where $h_{i}$ is the helicity of $i$th photon. The remaining amplitudes with opposite helicity
fermions can be obtained via parity symmetry.

These amplitudes were originally computed in  \cite{Kleiss:qed} and are given by
\begin{align}
	A_{n;q}^{\tree}(q^-,\qb^+;1^{h_1},\ldots,n^{h_n}) &=
	\frac{i}{\prod_{j=1}^{n} \la \xi_j  k_j \ra }\sum_{\sigma\in\mathfrak{S}_n}
	F_{n;q}(q^-,\qb^+;\sigma(1)^{h_{\sigma(1)}},\ldots,\sigma(n)^{h_{\sigma(n)}})\,,\\
	F_{n;q}(q^-,\qb^+;1^{h_1},\ldots,n^{h_n})&=\A{a_1}{q}\B{\qb}{b_n}\prod_{i=1}^{n-1}
	\frac{\AB{a_i}{q+K_{1,i}}{b_{i+1}}}{(q+K_{1,i})^2} \,,
	\label{eq:nphtntree}
\end{align}
where $\{\xi_k\}$ is a set of light-like reference vectors and
$K_{1,i} = \sum_{j=1}^{i} k_{i}$. We have also defined
\begin{align}
	&a_i={1+h_i\over2}\, \xi_{i} +{1-h_i\over2}\,k_i,    \qquad
	b_i={1+h_i\over2}\, k_i+{1-h_i\over2}\, \xi_{i} \,.
\end{align}
The definitions of for spinor products follow the standard conventions used in the QCD literature and are 
summarised in the appendix.

In the MHV case the amplitude one can show \cite{Ozeren:qed} 
that the amplitude takes the following simplified form,
\begin{equation}
	\label{eq:KSMHV}
	A^\tree_{n;q}\big(q^-,\qb^+;1^-,2^+,\dots,n^+\big)=
	i \frac{\A q\qb^{n-2} \A 1q^2}{\prod_{\alpha=2}^{n}\A q\alpha\A \qb\alpha}\,.
\end{equation}

\subsection{The $S\bar{S}+n(\gamma)$ amplitudes}

The tree-level amplitudes with a massive (complex) scalar coupling to an arbitrary number of photons
are given as \cite{Badger:2008rn}
\begin{equation}
  A^\tree_{n;S}\big(S,\bar{S};k_1,\dots,k_n\big)=i \,
\sum_{\sigma\in\mathfrak{S}_n}\,
F_{n;S}\big(S,\bar{S};k_{\sigma(1)},\dots,k_{\sigma(n)}\big)\,,
\end{equation}
of an amplitude defined from the partition of the $n$ ordered external
legs partitioned in group of at most length two
\begin{equation}
  F_{n;S}\big(S,\bar{S};k_{\sigma(1)},\dots,k_{\sigma(n)}\big)
= \sum_{a_1+\cdots+ a_r=n\atop
   a_k\in\{1, 2\}}\,  \prod_{s=1}^r \, {\epsilon_{\sigma(a_1+\cdots
     + a_{s-1}+1)}\cdot H(a_s)\over
     (p_a+\sum_{j=1}^{a_1+\cdots+a_s}k_{\sigma(j)})^2-\mu^2}\,,
     \label{eq:mscalartree}
\end{equation}
with
\begin{equation}
  H(a_s)=
  \begin{cases}
    q+\sum_{j=1}^{a_1+\cdots   +a_{s-1}}  k_{\sigma(j)}&  \textrm{if}\
    a_s=1\\
\epsilon_{\sigma(a_1+\cdots   +a_{s})}&\textrm{if}\ a_s=2\,.
  \end{cases}\,
\end{equation}
Because  of the  cubic and  quartic vertices,  this amplitude  is
a much larger sum of terms than the fermionic case. The number can be expressed as a sum over $n!\times
F_{n+1}$ where $F_r$ is the Fibonacci number of order $r$ (such that
$F_0=F_1=1$ and $F_2=2$).

\subsection{Large $z$ scaling}

We consider the polynomial behaviour of tree level amplitudes listed above under large values of the
complex parameter used in the BCFW recursion relations. We define such a complex momentum shift as
$\la a,b]$ where\footnote{Note that this definition differs from other instances in the literature
where $\la a,b]$ corresponds to a shift vector of $\AB{b}{\gamma^\mu}{a}$.},
\begin{align}
  \wh{a}^\mu &= a^\mu - \frac{z}{2}\AB{a}{\gamma^\mu}{b} \,, & 
  \wh{b}^\mu &= b^\mu + \frac{z}{2}\AB{a}{\gamma^\mu}{b} \,. & 
  \label{eq:BCFWshiftdef}
\end{align}
Recently it was shown that one-loop multi-photon amplitudes have a surprisingly simple
structure \cite{Badger:2008rn}. This can be explained through analysing the large $z$ behaviour of the tree level
amplitudes entering generalised cuts in the loop amplitude, a technique that has
successfully uncovered similar cancellations in gravity theories \cite{Bern:2007xj}. The key insight
of \cite{Badger:2008rn} was to demonstrate an improvement
in the behaviour of the $q\qb+$photon tree amplitudes under large values of $z$ when shifting
the fermion pair,
\begin{equation}
	A_{n;q}^{\tree}(q^-,\qb^+;1^{h_1},\ldots,n^{h_n}) \overset{\la q,\qb];z\to\infty}{\sim}
	\frac{C_\infty}{z^{n-1}} \,.
	\label{eq:qedzscale}
\end{equation}
This improved scaling only appears after the permutation sum has been performed and is independent
of the helicities of the photon lines. 

It was also observed that the amplitudes with a pair of massive scalars also share the same property,
\begin{equation}
	A_{n;S}^{\tree}(S^-,\Sb^+;1^{h_1},\ldots,n^{h_n}) \overset{\la \fl S,\fl\Sb];z\to\infty}{\sim}
	\frac{C_\infty}{z^{n-2}} \,.
	\label{eq:Szscale}
\end{equation}

\section{Dressing the BCFW relation}
The on-shell BCFW recursion relation can be derived by considering a complex contour integral over
the function $A(z)/z$:
\begin{equation}
	A_\infty = \frac{1}{2\pi i}\oint dz \frac{A(z)}{z} = A(0) - \sum_{\text{residues }z_k}
	A_L(z_k)\frac{i}{P_k^2}A_R(z_k) \,,
	\label{eq:bcfw}
\end{equation}
where $z_k={P_k^2}/{\AB{a}{P_k}{b} }$ for a $\la a,b]$ shift and the momentum $P_k$ flows from
right to left\footnote{The factor of $i$ in the propagator is specific to gluons and scalars.
Fermion propagators require and additional factor of $- i$ as we will see later.}.
the term $A_\infty$ is zero as long as $A(z)\overset{z\to\infty}{\to} \mathcal{O}({1}/{z})$ or better.
For the QED amplitudes considered above we observed that under certain shifts the large $z$
behaviour was much better than this minimum requirement. This allows us the freedom to consider a
new integral which will still evaluate to zero:
\begin{equation}
	0 = \frac{1}{2\pi i}\oint dz \frac{\alpha-z}{\alpha} \frac{A(z)}{z} = A(0) - \sum_{\text{residues }z_k}
	A_L(z_k) \frac{\alpha-z_k}{\alpha} \frac{i}{P_k^2}A_R(z_k) \,.
\end{equation}
Since we are free to choose $\alpha$ we can use this factor to cancel one of the poles in $A(z)$
and therefore reduce the number of terms in $A(0)$ compared to the representation of eq.
\eqref{eq:bcfw}. 
The fact that an improved large $z$ behaviour of the amplitudes can be used to derive
simplified expressions for tree-level amplitudes has been observed previously in \cite{ArkaniHamed:tree}.
For one inserted dressing factor our formula is essentially identical to the approach used in
\cite{Spradlin:bonus}
to derive relations between supergravity amplitudes.
For our QED amplitudes under the $\la q,\bar{q}]$ shift we have the modified boundary
behaviour of $z^{1-n}$ and so we can reduce the number of diagrams in the recursion
relation by introducing $(n-2)$ additional propagator factors. The recursion relation then takes the
following form:
\begin{align}
	0 = -\frac{1}{2\pi i}\oint \frac{dz}{z} A(z)\prod_{l=1}^{n-2}
	\frac{z_l -z}{z_l} = 
	A(0)
	- \sum_{\text{residues }z_k} A_L(z_k)\frac{i\mathcal{F}_n(P_k)}{P_k^2}A_R(z_k) \,,
	\label{eq:dressedbcfw}
\end{align}	
where 
\begin{equation}
	\mathcal{F}_n(P_k) = \prod_{l=1}^{n-2} \frac{z_l-z_k}{z_l}
	= \frac{1}{ {\AB{q}{P_k}{\bar{q}}}^{n-2} }  \prod_{l=1}^{n-2}
	\frac{\AB{q}{P_k(P_l-P_k)P_l}{\bar{q}}}{  P_{l}^2 } \,.
		\label{eq:dressingF}
\end{equation}
We see that the dressing factors contain all the explicitly removed propagators, $P_l^2$, thus ensuring that
the amplitude has the correct pole structure.
It is interesting to note that under a subsequent $\la q,\bar{q}]$ shift \eqref{eq:dressingF} 
falls off as $z^{2-n}$ due to these additional propagator factors.
Hence if we compare the expression for an amplitude computed from the dressed BCFW relation \eqref{eq:dressedbcfw} 
to that computed from the standard one \eqref{eq:bcfw}, we notice the following: apart from consisting of fewer terms
each term in the former expression will have the improved large $z$ behaviour.
So if it is known that an amplitude has a certain large $z$ behaviour,
this can be made manifest term by term using the dressed recursion relations.
In addition, the formula obtained in this way will consist of fewer terms compared to
a formula obtained from the standard recursion relations.

\subsection{Re-derivation of the Kleiss-Stirling MHV amplitude \label{sec:KS}}

\begin{figure}[t]
	\begin{center}
		\psfrag{a}{(a)}
		\psfrag{b}{(b)}
		\psfrag{1}{$1^-$}
		\psfrag{2}{$2^+$}
		\psfrag{k}{$k^+$}
		\psfrag{n}{$n^+$}
		\psfrag{F}{\hspace*{-2mm} $\mathcal{F}$}
		\psfrag{qb}{$\wh\qb^+$}
		\psfrag{q}{$\wh q^-$}
		\psfrag{sum}{
			\begin{minipage}{2cm}
				\begin{equation*}
					\sum_{k=3}^{n}	
				\end{equation*}
			\end{minipage}
		}
		\includegraphics[width=13cm]{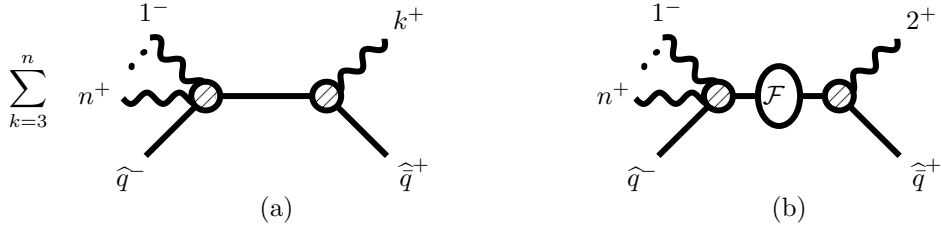}
	\end{center}
	\caption{(a) The $(n-1)$ diagrams contributing to the standard BCFW recursion relation for the
	$q\qb+n$-photon amplitudes. (b) The modified recursion containing only a single term.}
	\label{fig:MHVbcfw}
\end{figure}

In this section we re-derive the photon-MHV amplitude of eq. \eqref{eq:KSMHV}. 

We choose the $\la q,\bar{q}]$ shift and take the momentum of the negative helicity to photon to be
$p_1$. Since the amplitude is unordered this is done without loss of generality. We
choose the $(n-2)$ dressing factors such that we cancel contributions from the two-particle channels
with the anti-quark and the positive helicity photons $p_3, \ldots, p_n$. In the MHV case we must have at
least one negative helicity particle in each tree level sub-amplitude or we will get a vanishing
contribution. The only exception to this is the three-point \MHVb amplitude. There are therefore
$(n-1)$ contributing terms to the standard BCFW recursion relation for this helicity configuration as
shown in figure \ref{fig:MHVbcfw}. However the modified recursion relation has only a single
contribution from the $P_{\qb 2}$ channel 
(we denote $P_{\qb 2}^{\mu} =\qb^{\mu} + k_{2}^{\mu}$ and  $(P_{\qb 2})^2 = s_{\qb 2}$)
where the dressing factor is given by:
\begin{align}
	\mathcal{F}_n(P_{\qb 2})&=\prod_{k=3}^n \frac{\AB {q} {P_{\qb 2}(P_{\qb k}-P_{\qb 2})P_{\qb k}} {\qb} }
	{\AB{q}{P_{\qb 2}}{\qb}s_{\qb k}} \nonumber\\
	&= \frac{\A{q}{\qb}^{n-2}}{\A{q}{2}^{n-2}} \prod_{k=3}^n \frac{\A 2k}{\A{\qb}{k}}
	\label{eq:MHVdressingfactor}
\end{align}
The modified recursion relation then simply becomes:
\begin{align}\label{MHVrecursion}
	A^\tree_{n;q}(q^-,\qb^+;1^-,2^+,\dots,n^+) &= -i 
	A^\tree_{n-1;q}(\wh q^-,\wh{P}_{\qb 2}^+;1^-,3^+\ldots,n^+)
	\frac{i\mathcal{F}_n(P_{\qb 2})}{s_{\qb 2}}
	A^\tree_{1;q}(-\wh{P}_{\qb 2}^-,\wh\qb^+;2^+) 
	\,.
\end{align}
This can be solved inductively using the Kleiss-Stirling formula \eqref{eq:KSMHV} as an ansatz. 
In order to prove \eqref{eq:KSMHV} recursively we need the following two three-point amplitudes,
\begin{align}
A^\tree_{1;q}\big(q^-,\qb^+;1^-\big)=
	i \frac{  \A 1q^2}{\A q\qb}\,,
\end{align}
 and 
\begin{align}
\tilde{A}^\tree_{1;q}\big(q^-,\qb^+;1^+\big)=
	i \frac{  [ 1\qb]^2}{ [q\qb]}\,,
\end{align} 
which follow from the vertices in the QED Lagrangian.
Notice that the first formula agrees with \eqref{eq:KSMHV} for $n=1$.
It is then sufficient to assume that \eqref{eq:KSMHV}
holds for $(n-1)$ photons and show that it holds for $n$ photons.
Indeed, using
\begin{align}
A^\tree_{n-1;q}(\wh q^-,\wh{P}_{\qb 2}^+;1^-,3^+,..,n^+) =  i \frac{ {\A q 2 }^{n-3} {\A 1 q}^2 [ \wh{P}_{\qb 2}  \qb] }{[2 \qb ] \, \prod_{k=3}^{n} {\A q k} {\A 2 k } }\,,
\end{align}
and (see (\ref{minusbrackets}))
\begin{align}
 A^\tree_{1;q}(-\wh{P}_{\qb 2}^-,\wh\qb^+;2^+) = \frac{ [2 \qb ]^2 }{[\wh{P}_{\qb 2} \qb]} \,,
\end{align}
and (\ref{eq:MHVdressingfactor})
and plugging them into (\ref{MHVrecursion}) we immediately obtain
\begin{align}
	A^\tree_{n;q}(q^-,\qb^+;1^-,2^+,\ldots,n^+) 
	&= i\frac{\A q1^2}{\A q\qb} \prod_{k=2}^n \frac{\A q\qb}{\A q k \A \qb k}\,.
\end{align}
This shows us that the modified recursion relations give an efficient way to derive compact
expressions for the QED amplitudes. The simplifications from the permutation sums are essentially
factored into the $\mathcal{F}$ functions modifying the propagator.

\subsection{NMHV amplitudes \label{sec:NMHV}}

\begin{figure}[t]
	\begin{center}
		\psfrag{1}{$1^-$}
		\psfrag{2}{$2^-$}
		\psfrag{k}{$k^+$}
		\psfrag{n}{$n^+$}
		\psfrag{F}{$\mathcal{F}$}
		\psfrag{N}{\bf N}
		\psfrag{qb}{$\wh\qb^+$}
		\psfrag{q}{$\wh q^-$}
		\psfrag{sum1}{
			\begin{minipage}{2cm}
				\begin{equation*}
					\sum_{k=3}^{n}	
				\end{equation*}
			\end{minipage}
		}
		\psfrag{sum2}{\hspace{-7mm}
			\begin{minipage}{2cm}
				\begin{equation*}
					+\,\sum_{\rm partitions}
				\end{equation*}
			\end{minipage}
		}
		\psfrag{N}{\tiny\rm\bf N}
		\psfrag{(1-2)}{\hspace{-7mm}$+(1\leftrightarrow2)$}
		\includegraphics[width=0.9\textwidth]{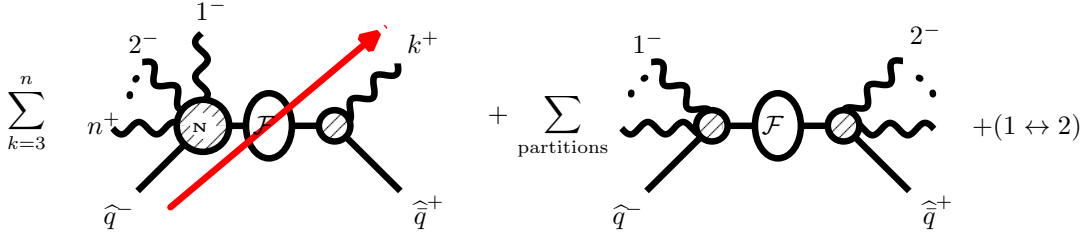}
	\end{center}
	\caption{Of the two topologies contributing to the NMHV photon amplitudes all of the
	diagrams involving lower point NMHV vertices can be eliminated by dressing the recursion relation}
	\label{fig:NMHV}
\end{figure}

We now turn our attention to the NMHV amplitudes with two negative helicity photons. The first
non-trivial of these occurs for $n=4$ and a compact form for it has previously been derived using
a standard BCFW shift of the anti-fermion and a negative helicity photon, $\la 1,\qb]$, \cite{Ozeren:qed}: 
\footnote{In order not to make the notation too heavy, in the following we will abbreviate e.g. $p_{j}$ by $j$.
We hope that this does not lead to confusion.} 
\begin{align}
	\label{eq:OS_NMHV}
	A^\tree_{4;q}(q^-,\qb^+;1^-,2^-,3^+,4^+) =& 
	P(1^-,2^-,3^+,4^+)+P(1^-,2^-,4^+,3^+)\nonumber\\
	+Q(1^-,2^-,3^+,4^+)+&Q(1^-,2^-,4^+,3^+)
	+R(1^-,2^-,4^+,3^+)\,,
\end{align}
where
\begin{align}
	P(q^-,\qb^+;1^-,2^-,3^+,4^+) &= 
		\frac{\AB{1}{\qb+3}{q}\AB{1}{\qb+3}{4}^2}
		{s_{\qb 13}\B q2\A \qb 3\AB{3}{\qb+1}{q}\AB{1}{\qb+p3}{2}} \,, \\
	Q(q^-,\qb^+;1^-,2^-,3^+,4^+) &=
		\frac{\A q1^2\B 3\qb^2\AB{1}{3+2}{\qb}}
		{s_{\qb 23}\A q4\B 1\qb\AB{4}{3+2}{\qb}\AB{1}{\qb+3}{2}}\,, \\
	R(q^-,\qb^+;1^-,2^-,3^+,4^+) &= 
		\frac{s_{234}\B q\qb^2\AB{2}{q+1}{\qb}^2}
		{\AB{3}{q+1}{\qb}\AB{4}{q+1}{\qb}\AB{3}{\qb+1}{q}\AB{4}{\qb+1}{q}\B1\qb\B q1}\,.
\end{align}
An observation one can make 
is that the functions $P$ and $Q$ both scale as ${1}/{z^2}$ rather than
the ${1}/{z^3}$ behaviour of the full amplitude in the large $z$ limit. Using the dressed
recursion relation and a $\la q,\qb]$ shift we can derive an alternative formula in which each term
scales as ${1}/{z^3}$.

We begin by choosing the $(n-2)$ diagrams with a three-point \MHVb-amplitude to vanish by using the same
dressing factor as in the MHV case,
\begin{equation}
	\mathcal{F}_n(P) = \prod_{l=3}^{n} \frac{z_l-z_P}{z_l} =  \frac{1}{\AB{q}{P}{\qb}^{n-2}}\prod_{l=3}^{n}
	\frac{\AA{q}{P(\qb-P)}{l}}{\A{\qb}{l}} \,.
	\label{eq:dressNMHV}
\end{equation}
The recursion relation then takes the form of a sum over products of two MHV amplitudes:
\begin{align}
	A^\tree_{n;q}(q^-,\qb^+,1^-,2^-,3^+,\ldots,n^+) & 
	=\nonumber\\
	& \hspace{-5cm} = 
	 \sum_{\sigma\in S_2}\sum_{P_1|P_2}
	-iA^\tree_{n_1+1;q}(\wh{q}^-,\wh{Q_1}^+,1^-,\{P_1\})
	\frac{i\mathcal{F}_n(Q_1)}{Q_1^2}
	A^\tree_{n_2+1;q}(-\wh{Q_1}^-,\wh{\qb}^+,2^-,\{P_2\})\nonumber\\
	& \hspace{-5cm} \equiv  \sum_{\sigma\in S_2}\sum_{P_1|P_2}
	F_{n_1,n_2}(q^-,\qb^+,1^-,2^-,\{P_1\},\{P_2\})\,.
\end{align}
There is only a single topology left for the full NMHV amplitude as shown in figure \ref{fig:NMHV}:
\begin{align}
	F_{n_1,n_2}(q^-,\qb^+,1^-,2^-,\{P_1\},\{P_2\}) &= A^\tree_{n_1+1;q}(\wh{q}^-,\wh{Q_1}^+,1^-,\{P_1\})\frac{\cF_n(Q_1)}{Q_1^2}A^\tree_{n_2+1;q}(-\wh{Q_1}^-,\wh{\qb}^+,2^-,\{P_2\}) \,,
\end{align}
where
\begin{equation}
	Q_1 = \qb+2+\sum_{k\in P_2} k \,.
\end{equation}
We also define $n_1$ and $n_2$ to be the number of positive helicity photons in the left and right amplitudes. Using
the expressions for the MHV amplitudes and expanding the shifted spinors we find,
\begin{align}\label{eq:FNMHV}
F_{n_1,n_2}(q^-,\qb^+,1^-,2^-,\{P_1\},\{P_2\}) =&\nonumber\\
& \hspace{-4cm}  \frac{
	-i\A q1^2\AB{2}{Q_1}{\qb}^2
}{
	Q_1^2(Q_1-\qb)^2\AB{q}{Q_1}{\qb}
}	
\prod_{k\in P_1}
\frac{
	\AA{q}{Q_1(Q_1-\qb)}{k}
}{
	\A qk \A k \qb \AB{k}{Q_1}{\qb}
}
\prod_{k\in P_2}
\frac{
	(Q_1-\qb)^2
}{
	\A \qb k \AB{k}{Q_1}{\qb}
} \,.
\end{align}
Restricting ourselves to the $n=4$ NMHV amplitude we can explicitly write
\begin{align}
	\label{eq:NMHV}
	A^\tree_{4;q}(q^-,\qb^+;1^-,2^-,3^+,4^+) =&  \nonumber\\
	&  \hspace{-4cm}  F^1_4(1^-,2^-,3^+,4^+)+F^1_4(2^-,1^-,3^+,4^+) + F^2_4(1^-,2^-,3^+,4^+) + \nonumber \\
	& \hspace{-4cm}  F^2_4(1^-,2^-,4^+,3^+)+F^2_4(2^-,1^-,3^+,4^+)+F^2_4(2^-,1^-,4^+,3^+)\,,
\end{align}
where
\begin{align}	
	&F^{1}_{4}(1^-,2^-,3^+,4^+)=
	\frac{i s_{234} \la 2|3+4|\qb]^2}
	{\la 3\qb\ra  \la 4\qb\ra  \la 3|2+4|\qb] \la 4|2+3|\qb][1q] [1\qb]} \,,\\
	&F^{2}_{4}(1^-,2^-,3^+,4^+)=
		\frac{i \la q1\ra ^2 \la 24\ra  \la q|1+3|2+4|3\ra  [\qb4]^2}
		{s_{q13} \la 3q\ra  \la 3\qb\ra  \la 4\qb\ra  \la q|2+4|\qb] \la 3|2+4|\qb] [\qb2]}\,,
\end{align}
and where it is simple to check that eq. \eqref{eq:NMHV} and \eqref{eq:OS_NMHV} agree numerically.

Comparing these two equations one notices that our new
improved
recursion relation yields a representation which is actually one term longer than the
standard BCFW $\la\gamma,\qb]$ shift. It turns out that this is not a general feature and it is the
purpose of the next section to show that in general \eqref{eq:FNMHV} is considerably more compact 
compared to the previously
known results.

\subsection{Number of terms in the amplitudes}
\label{sect:nmhv-counting}

Let us compare the number of terms for the NMHV amplitudes considered above, 
and compare between the standard and dressed recursion relation.

Recall that $n$ is the total number of photons. We denote the number of plus helicity photons
by $m$. For the NMHV amplitude, $m=n-2$.
For a BCFW shift which involves the fermion/anti-fermion pair we get the following recurrence
relation for the number of terms $f^{\rm NMHV}_{m}$ in the NMHV amplitude,
\begin{align}
f^{\rm NMHV}_m= m f^{\rm NMHV}_{m-1} +2 g_m\,,
\end{align}
with $g_m=\sum_{i=1}^{m}{m!}/(m-i)! / i! = 2^{m}-1$  and $f^{\rm NMHV}_0=0$.
Here the factor $m$ comes from the ways of choosing the plus helicity photon on the $\overline{\rm MHV}_{3}$ vertex,
and $g_{m}$ counts the number of diagrams built from two MHV vertices. The factor two accounts for the two
possible positions of the negative helicity photons.
The solution to this recurrence is
$f^{\rm NMHV}_m = 2 m!   \left( \sum_{k=0}^{m} {2^k}/{k!} - \sum_{k=0}^{m}{1}/{k!} \right)$, 
so $f^{\rm NMHV}_m$ grows factorially for large $m$, 
\footnote{It is possible to obtain slightly improved expressions, 
but which still exhibit a factorial growth.
For example, shifting the anti-fermion and one of the negative 
helicity photons as in  \cite{Ozeren:qed}, one obtains $f^{\rm NMHV}_m \sim (e^2-e) m!$.
Also, by making a convenient choice of the reference spinors in the
Kleiss-Stirling formula \eqref{eq:nphtntree} one finds $f^{\rm NMHV}_m = (m-1)!$ \cite{Badger:2008rn}.
}
\begin{align}
f^{\rm NMHV}_{m} \sim 2 (e^2-e) m!\,.
\end{align}

Let us now study the improvement induced by the dressed recursion relations discussed in section
\ref{sec:NMHV}.
Recall that thanks to the $z^{-m-1}$ falloff for large $z$ we can introduce $m$ dressing factors 
in such a way that they eliminate the $m$ diagrams homogeneous in the NMHV amplitude.
Hence we will obtain an improved relation for the number of terms in the amplitude,
\begin{equation}
{f}^{\rm NMHV,dressed}_m = 2 g_m = 2 (2^m -1)  \sim 2^{m+1}\,,
\end{equation}
which grows {\it exponentially} instead of {\it factorially}.
While for small values of $m$ one finds a comparable number of terms the advantage of the dressed
recursion relations becomes obvious for bigger values of $m$.
Some sample values are given in the following table for illustration.
\begin{table}[t] 
\caption{Number of terms obtained from conventional vs dressed BCFW recursion.
$m$ is the number of plus helicity photons.}
\begin{center}
\begin{tabular}{|c|c|c|c|c|c|c|c|c|c|c| }
\hline
$m$  & 1 & 2 &3 &4 &5 &6 &7 &8 &9 & 10 \\
\hline
$f^{\rm NMHV}_{m}/2$ & 1 & 5 & 22 & 103 & 546 & 3339 & 23500 & 188255 & 1694806 & 16949083\\
\hline
$f^{\rm NMHV,dressed}_{m}$ & 2 & 6 & 14 & 30 & 62 & 126 & 254 & 510 & 1022 & 2046\\ 
\hline
\end{tabular}
\end{center}
\label{default}
\end{table}%

\subsection{N$^2$MHV amplitudes}

\begin{figure}[t]
  \begin{center}
	\psfrag{1}{$1^-$}
	\psfrag{2}{$2^-$}
	\psfrag{3}{$3^-$}
	\psfrag{P1}{$P_1$}
	\psfrag{P2}{$P_2$}
	\psfrag{P3}{$P_3$}
	\psfrag{k}{$k^+$}
	\psfrag{n}{$n^+$}
	\psfrag{N}{\small\bf N}
	\psfrag{F}{$\mathcal{F}$}
	\psfrag{qb}{$\wh\qb^+$}
	\psfrag{q}{$\wh q^-$}
	\psfrag{a}{(a)}
	\psfrag{b}{(b)}
	\psfrag{c}{(c)}
    \includegraphics[width=14cm]{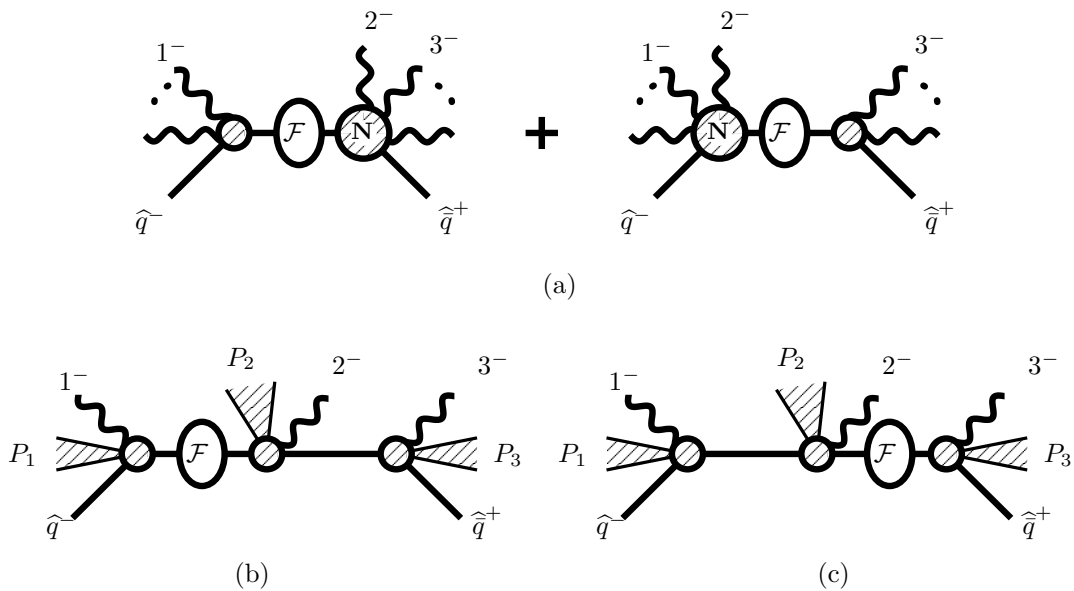}
  \end{center}
  \caption{(a) N$^2$MHV amplitude as the sum over products of MHV and NMHV amplitudes in the dressed
  recursion relation. (b) Diagrammatic representation of the function $G$. (c) Diagrammatic
  representation of the function $G'$.}
  \label{fig:NNMHV}
\end{figure}

In this section we derive closed form analytic expressions for the N$^2$MHV amplitudes with three
negative helicity photons using the dressed recursion relation. Just as in the NMHV case we can use
the dressing factor to remove all homogeneous diagrams so that the NMHV amplitude is simply a
product of NMHV and MHV amplitudes as shown in figure \ref{fig:NNMHV}(a).
We write the dressed recursion relation as:
\begin{align}
	&A^\tree_{n;q}(q^-,\qb^+,1^-,2^-,3^-,4^+,\ldots,n^+)
	=\nonumber\\
	=&
	\sum_{\sigma\in S_3/\mathbb{Z}}\sum_{P_L|P_R}
	-iA^\tree_{n_L;q}(\wh{q}^-,\wh{Q_1}^+,\sigma(1)^-,\{P_L\})
	\frac{i\mathcal{F}_n(Q_1)}{Q_1^2}
	A^\tree_{n_R;q}(-\wh{Q_1}^-,\wh{\qb}^+,\sigma(2)^-,\sigma(3)^-,\{P_R\})
	\nonumber\\
	&+
	\sum_{\sigma\in S_3/\mathbb{Z}}\sum_{P_L|P_R}
	A^\tree_{n_L;q}(\wh{q}^-,\wh{Q_1}^+,\sigma(1)^-,\sigma(2)^-,\{P_L\})
	\frac{\mathcal{F}_n(Q_1)}{Q_1^2}
	A^\tree_{n_R;q}(-\wh{Q_1}^-,\wh{qb}^+,\sigma(3)^-,\{P_R\})
	\nonumber\\
	=&
	\sum_{\sigma\in S_3}\Bigg(	
	\sum_{P_1|P_2|P_3}
	G_{n_1,n_2,n_3}(q^-,\qb^+,\sigma(1)^-,\sigma(2)^-,\sigma(3)^-,\{P_1\},\{P_2\},\{P_3\})
	\nonumber\\&\hspace{5mm}
	+\sum_{P_1'|P_2'|P_3'}
	G'_{n_1,n_2,n_3}(q^-,\qb^+,\sigma(1)^-,\sigma(2)^-,\sigma(3)^-,\{P_1'\},\{P_2'\},\{P_3'\})
	\Bigg).
\end{align}
Each $G,G'$ function represents a single term in the final result expressed as a product of an MHV
amplitude with the NMHV $F$ function defined in equation (\ref{eq:FNMHV}),
\begin{align}
	&G_{n_1,n_2,n_3}(q^-,\qb^+,1^-,2^-,3^-,\{P_1\},\{P_2\},\{P_3\})
	=\nonumber\\&\hspace{1cm}
	A^\tree_{n_1+1;q}(\wh{q}^-,\wh{Q_1}^+;1^-,\{P_1\})
	\frac{\mathcal{F}_n(Q_1)}{Q_1^2}
	F_{n_2,n_3}(-\wh{Q_1}^-,\qb^+,2^-,3^-,\{P_2\},\{P_3\})\,,
\end{align}
and
\begin{align}
	&G'_{n_1,n_2,n_3}(q^-,\qb^+,1^-,2^-,3^-,\{P_1\},\{P_2\},\{P_3\})
	=\nonumber\\&\hspace{1cm}
	F_{n_1,n_2}(\wh{q}^-,\wh{Q_2}^+,1^-,2^-,\{P_1\},\{P_2\})
	\frac{\mathcal{F}_n(Q_2)}{Q_2^2}
	A^\tree_{n_3+1;q}(-\wh{Q_2}^-,\qb^+;3^-,\{P_3\})\,,
\end{align}
where
\begin{align}
	Q_1 &= \qb+2+3+\sum_{k\in P_2\cup P_3} k\,, &
	Q_2 &= \qb+3+\sum_{k\in P_3} k \,.
\end{align}
These topologies are written graphically in figure \ref{fig:NNMHV}(b) and (c).
These functions can be quickly written in closed form:
\begin{align}
&G_{n_1,n_2,n_3}(q^-,\qb^+,1^-,2^-,3^-,\{P_1\},\{P_2\},\{P_3\})=\nonumber\\&
\frac{
	i\A q1^2\AB{2}{Q_1}{\qb}^2\AB{3}{Q_2}{\qb}^2\AA{q}{Q_1(Q_1-\qb)}{3}
}{
	Q_1^2(Q_1-Q_2)^2\AB{q}{Q_1(Q_2-Q_1)Q_2}{\qb}\BB{\qb}{Q_1Q_2}{\qb}\AB{q}{Q_1}{\qb}\A{\qb}{3}
}\nonumber\\&
\prod_{k\in P_1}
\frac{
	\AA{q}{Q_1(Q_1-\qb)}{k}
}{
	\A qk\A{\qb}{k}\AB{k}{Q_1}{\qb}
}
\prod_{k\in P_2}
\frac{
	\AB{k}{(Q_2-\qb)(Q_1-Q_2)Q_1}{\qb}
}{
	\A{k}{\qb}\AB{k}{Q_2}{\qb}\AB{k}{Q_1}{\qb}
}
\prod_{k\in P_3}
\frac{
	(Q_1-Q_2)^2
}{
	\A{k}{\qb}\AB{k}{Q_2}{\qb}\AB{k}{Q_1}{\qb} 
} \,,
\end{align}
and
\begin{align}
&G'_{n_1,n_2,n_3}(q^-,\qb^+,1^-,2^-,3^-,\{P_1\},\{P_2\},\{P_3\})=\nonumber\\&
\frac{
	i\A q1^2\AA{2}{(Q_1-Q_2)Q_1}{q}^2\AB{3}{Q_2}{\qb}^2\AA{q}{Q_2(Q_2-\qb)}{3}
}{
	Q_2^2(Q_2-Q_1)^2(Q_2-\qb)^2\AA{q}{Q_2Q_1}{q}\AB{q}{Q_2}{\qb}\A{\qb}{3}
}
\prod_{k\in P_1}
\frac{
	\AA{q}{Q_1(Q_1-Q_2)}{k}\AA{q}{Q_2(Q_2-\qb)}{k}
}{
	\A{q}{k}\AB{k}{Q_2}{\qb}\AA{k}{(Q_1-Q_2)Q_2}{q}\A{\qb}{k}
}\nonumber\\&
\prod_{k\in P_2}
\frac{
	(Q_1-Q_2)^2\AA{q}{Q_2(Q_2-\qb)}{k}
}{
	\AB{k}{Q_2}{\qb}\AA{k}{(Q_1-Q_2)Q_2}{q}\A{\qb}{k}
}
\prod_{k\in P_3}
\frac{
	(Q_2-\qb)^2
}{
	\AB{k}{Q_2}{\qb}\A{k}{\qb}
} \,.
\end{align}
The full N$^2$MHV amplitude for $n=6$ contains $186$ terms in the dressed case versus $720$ for the Feynman diagram
computation. 
The counting of terms can be done as in section \ref{sect:nmhv-counting}. For the N$^2$MHV
amplitude obtained form the dressed recursion we arrive at a total number of terms ($m$ is the number
of plus helicity photons, i.e. $m=n-3$ for N$^2$MHV):
\begin{align}
f^{\rm N^2MHV}(m) = 12 \times 3^m - 18  \times 2^m + 6\,.
\end{align}
We chose to use the same dressing factor as in the NMHV case even though 
here the $z=z_{3\qb}$ channel (corresponding to $l=3$ in eq. \eqref{eq:dressNMHV}) vanishes
explicitly. Although one could choose it to be another non-zero diagram
the saving in complexity would
be modest and at the cost of losing the symmetry of the final answer.

\subsection{Massive scalar amplitudes \label{sec:mscalar}}

In this section we consider amplitudes with photons and a pair for massive
(complex) scalars. The amplitudes with gluons have been computed using a massive BCFW
recursion in reference \cite{Badger:massrec}. We can obtain the two-photon amplitude by summing over
permutations of the gluon result:
\begin{align}
  A_{4;S}^{\tree}(S^+,1^+,2^+,\Sb^-) &=
  A_{4;S}^{\tree}(S^+,1_g^+,2_g^+,\Sb^-)
  +A_{4;S}^{\tree}(S^+,2_g^+,1_g^+,\Sb^-) \nonumber\\&=
  \frac{im^2\B12}{\A12\AB1S1}
  +\frac{im^2\B12}{\A12\AB2S2}.
\end{align}
Here we derive this four-point amplitude using the dressed recursion and shifting the two massive
particles as described by Schwinn and Weinzierl \cite{Schwinn:massrec}.
For this we first define a
basis of two massless vectors from the original massive pair:
\begin{align}
	\fl S &= \frac{\gamma(\gamma S+m^2\Sb)}{\gamma^2-m^4} &
	\fl \Sb &= \frac{\gamma(\gamma \Sb+m^2S)}{\gamma^2-m^4} &
	\gamma &= S\cdot\Sb + \sqrt{(S\cdot\Sb)^2-m^4}.
\end{align}
We first compute the all-plus configuration which vanishes in the massless limit. This amplitude
actually has a further improved boundary behaviour with respect to the universal scaling and goes as
${1}/{z^n}$ in the large $z$ limit. We first define the dressing function which we will use for
the $n$-point function:
\begin{equation}
	\mathcal{F}_n(z) = \prod_{l=2}^n \frac{z_l-z}{z_l}\,,
\end{equation}
where
\begin{equation}
	z_l = \frac{\gamma\AB{l}{\Sb}{l}}{(\gamma-m^2)\AB{\fl S}{l}{\fl \Sb}}\,.
\end{equation}
The $n=2$ amplitude can then be represented as:
\begin{align}
	A^\tree_{4;S}(S,\Sb;1^+,2^+) =& A^\tree_{3;S}(\wh{S},\wh{P}_{\Sb 1};2^+)
	\frac{i\mathcal{F}_2(z_1)}{\AB{1}{\Sb}{1}}A^\tree_{3;S}(-\wh{P}_{\Sb 1},\wh{\Sb};1^+)\\
	=& \frac{im^2\mathcal{F}_2(z_1)\A{\fl S}{\fl\Sb}^2\B{\fl\Sb}{2}\B{\fl\Sb}{1}}{\gamma\AB{1}{\Sb}{1}\A{\fl S}{1}\A{\fl S}{2}} \,.
\end{align}
Using the momentum conservation for the flatted vectors, $(1+\frac{m^2}{\gamma})(\fl S + \fl
\Sb)+1+2=0$, we find:
\begin{align}
	\mathcal{F}_2(z_1) &= -\frac{s_{12}}{\AB 2\Sb2} \,, &
	\frac{\AB{\fl S}{\fl\Sb}{2}}{\AB{1}{\fl S}{\fl \Sb}} &= \frac{\A{\fl S}{1}\B 12}{\A{1}{2}\B
	2{\fl \Sb}} \,.
\end{align}
This allows to eliminate $S$ and $\Sb$ from the amplitude leaving,
\begin{equation}
	A^\tree_{4;S}(S,\Sb;1^+,2^+) = \frac{im^2\B12^2}{\AB{1}{S}{1}\AB{2}{S}{2}} \,,
\end{equation}
which matches the standard result. Higher multiplicity amplitudes for $n=5,6$ have been checked
numerically against eq. \eqref{eq:mscalartree} to verify the validity of the dressed recursion.
Since there is no simple choice of dressing factors as in the fermion amplitudes an all multiplicity
solution to the recursion is more difficult to obtain and we refrain from further generalisations for the
time being.

\section{Conclusions}

In this letter we have constructed a dressed version of the BCFW recursion relation which allows the
computation of compact analytic formula for tree-level amplitudes in QED. The construction relies on
the improved boundary scaling property first observed in \cite{Badger:2008rn} and can be applied to
any situation where the amplitude falls off as ${1}/{z^2}$ at the boundary of the integration
contour. The new NMHV and N$^2$MHV amplitude representations are shown to have an exponential rather than factorial growth
in the number of terms compared to the standard on-shell recursion.

Since all formulae have the improved scaling behaviour manifest they are much better suited to find
the cancellations in loop amplitudes explicitly. They would be particularly useful in finding closed
form expression for the multi-photon amplitudes at one-loop for which the current limit is the
eight-point MHV amplitude.

\section*{Acknowledgiments}
We are grateful to J.~Drummond for stressing that an improved large $z$ behaviour implies
relations between different BCFW terms, which initiated this project. 
We also thank P.~Benincasa for interesting discussions and P.~Vanhove for useful comments on the
manuscript. SB acknowledges support from the Helmholtz Gemeinschaft under contract VH-NG-105.

\appendix

\section{Conventions}

We use the standard QCD conventions for the two-component spinor helicity formalism.
\begin{align}
(p_1 + p_2 )^2 = 2 p_1 \cdot p_2 = \la 1 2 \ra [ 2 1 ]\,.
\end{align}
Here
\begin{align}
\la 1 2 \ra = \lambda_{1}^{\alpha} \lambda_{2 \alpha}\,,\qquad  [ 2 1 ] = \tilde{\lambda}_{2 \dot{\alpha}} \tilde{\lambda}_{1}^{\dot{\alpha}} \,.
\end{align}
The extended spinor product is defined as,
\begin{align}
\AB{q}{P_k}{\bar{q}}  = \lambda_{q}^{\alpha} P_{k \alpha \dot{\alpha}} \tilde{\lambda}_{\bar{q}}^{\dot{\alpha}}\,,
\end{align}
and
\begin{align}
P_{k}^2 = P_{k} \cdot P_{k} = \frac{1}{2} P_{k}^{\alpha \dot{\alpha}} P_{k {\alpha \dot{\alpha}} }\,.
\end{align}
When encountering negative momenta in the spinors appearing in the recursion relations we define:
\begin{align}\label{minusbrackets}
|-p] &= i |p] & |-p\ra &= i |p\ra \,.
\end{align}

\bibliographystyle{h-elsevier}
\bibliography{qedtrees}

\begin{thebibliography}{10}

\bibitem{Bern:1994zx}
Z. Bern et~al.,
\newblock Nucl. Phys. B425 (1994) 217, hep-ph/9403226.

\bibitem{Bern:2007dw}
Z. Bern, L.J. Dixon and D.A. Kosower,
\newblock Annals Phys. 322 (2007) 1587, 0704.2798.

\bibitem{Britto:2004ap}
R. Britto, F. Cachazo and B. Feng,
\newblock Nucl. Phys. B715 (2005) 499, hep-th/0412308.

\bibitem{Britto:2005fq}
R. Britto et~al.,
\newblock Phys. Rev. Lett. 94 (2005) 181602, hep-th/0501052.

\bibitem{Drummond:2008vq}
J.M. Drummond et~al.,
\newblock Nucl. Phys. B828 (2010) 317, 0807.1095.

\bibitem{Drummond:2009fd}
J.M. Drummond, J.M. Henn and J. Plefka,
\newblock JHEP 05 (2009) 046, 0902.2987.

\bibitem{Bern:2002kj}
Z. Bern,
\newblock Living Rev. Rel. 5 (2002) 5, gr-qc/0206071.

\bibitem{Dixon:2010gz}
L.J. Dixon,
\newblock (2010), 1005.2703.

\bibitem{Bern:2008qj}
Z. Bern, J.J.M. Carrasco and H. Johansson,
\newblock Phys. Rev. D78 (2008) 085011, 0805.3993.

\bibitem{Bern:2010ue}
Z. Bern, J.J.M. Carrasco and H. Johansson,
\newblock (2010), 1004.0476.

\bibitem{Bern:2010yg}
Z. Bern et~al.,
\newblock (2010), 1004.0693.

\bibitem{BjerrumBohr:2010zs}
N.E.J. Bjerrum-Bohr et~al.,
\newblock (2010), 1003.2403.

\bibitem{Tye:2010dd}
H. Tye and Y. Zhang,
\newblock (2010), 1003.1732.

\bibitem{ArkaniHamed:tree}
N. Arkani-Hamed and J. Kaplan,
\newblock JHEP 04 (2008) 076, 0801.2385.

\bibitem{Bianchi:2008pu}
M. Bianchi, H. Elvang and D.Z. Freedman,
\newblock JHEP 09 (2008) 063, 0805.0757.

\bibitem{Drummond:2009ge}
J.M. Drummond et~al.,
\newblock Phys. Rev. D79 (2009) 105018, 0901.2363.

\bibitem{Drummond:2008cr}
J.M. Drummond and J.M. Henn,
\newblock JHEP 04 (2009) 018, 0808.2475.

\bibitem{Bern:2007xj}
Z. Bern et~al.,
\newblock Phys. Rev. D77 (2008) 025010, 0707.1035.

\bibitem{ArkaniHamed:2008gz}
N. Arkani-Hamed, F. Cachazo and J. Kaplan,
\newblock (2008), 0808.1446.

\bibitem{BjerrumBohr:2008ji}
N.E.J. Bjerrum-Bohr and P. Vanhove,
\newblock JHEP 10 (2008) 006, 0805.3682.

\bibitem{Badger:2008rn}
S. Badger, N.E.J. Bjerrum-Bohr and P. Vanhove,
\newblock JHEP 02 (2009) 038, 0811.3405.

\bibitem{Spradlin:bonus}
M. Spradlin, A. Volovich and C. Wen,
\newblock Phys. Lett. B674 (2009) 69, 0812.4767.

\bibitem{Feng:2010ku}
B. Feng and C.Y. Liu,
\newblock (2010), 1004.1282.

\bibitem{Feng:2010my}
B. Feng, R. Huang and Y. Jia,
\newblock (2010), 1004.3417.

\bibitem{Kleiss:qed}
R. Kleiss and W.J. Stirling,
\newblock Phys. Lett. B179 (1986) 159.

\bibitem{Ozeren:qed}
K.J. Ozeren and W.J. Stirling,
\newblock JHEP 11 (2005) 016, hep-th/0509063.

\bibitem{Badger:massrec}
S.D. Badger et~al.,
\newblock JHEP 07 (2005) 025, hep-th/0504159.

\bibitem{Schwinn:massrec}
C. Schwinn and S. Weinzierl,
\newblock JHEP 04 (2007) 072, hep-ph/0703021.

\end{thebibliography}

\end{document}